\documentclass{ws-procs975x65}
\usepackage{amssymb,graphicx}

\def\be{\begin{equation}}
\def\ee{\end{equation}}
\def\bea{\begin{eqnarray}}
\def\eea{\end{eqnarray}}
\def\exp{\mathrm{exp}}
\def\ex{\mathrm{e}}

\def\susix{\mathrm{SU(6)}}
\def\suthr{\mathrm{SU(3)}}
\def\sutwo{\mathrm{SU(2)}}
\def\uone{\mathrm{U(1)}}
\def\sz{{S^1}/{Z_2}}

\begin{document}

\title{\bf \large Proton decay in 5D$-\susix$ GUT with orbifold $\sz$ breaking in Scherk-Schwarz mechanism}

\author{A. Hartanto$^{a}$, F.P. Zen$^{a,b}$, J.S. Kosasih$^{a,b}$ and L.T. Handoko$^{c,d}$}

\address{$^{a)}$Theoretical Physics Laboratory, THEPI, Faculty of Mathematics and Natural Sciences,
Institut Teknologi Bandung, Jl. Ganesha 10 Bandung 40135, Indonesia}

\address{$^{b)}$Indonesia Center for Theoretical and Mathematical Physics, Jl. Ganesha 10, Bandung 40132, Indonesia}

\address{$^{c)}$Group for Theoretical and Computational Physics, Research
Center for Physics, Indonesian Institute of Sciences, Kompleks
Puspiptek Serpong, Tangerang 15310, Indonesia}

\address{$^{d)}$Department of Physics, University of Indonesia,
Kampus UI Depok, Depok 16424, Indonesia }

\begin{abstract}
Proton decay within  5-dimensional $\susix$ GUT with orbifold $\sz$ breaking is investigated using Scherk-Schwarz mechanism. It is shown that in the model neither leptoquark like heavy gauge bosons nor violation of baryon number conservation are allowed due to the orbifold breaking parity splitting. These results prevent too short proton lifetime within the model.
\end{abstract}

\keywords{proton decay, $\susix$, GUT, symmetry breaking, extra dimension}

\bodymatter

\section{Introduction}
\label{sec:intro}

Despite the Standard Model (SM) of particle physics containing the electroweak $\sutwo_L \otimes \uone_Y$ and the strong $\suthr_C$ theories are in impressive agreements with most of experimental observables \cite{pdb}, both theories are not yet unified in a single  symmetry. The SM with symmetry $\suthr_C \otimes \sutwo_L \otimes \uone_Y$, is lacking of explaining the unification of three gauge couplings at a particular scale assuming  that our nature should be explained by a single unified theory, the so-called grand unified theory (GUT).

There are many types of GUT models, but most of them deploys larger symmetry than $\suthr_C \otimes \sutwo_L \otimes \uone_Y$ to accommodate at least all three interactions at the particle elementer scale. One of models containing $\suthr_C \otimes \sutwo_L \otimes \uone_Y$ as a part of its subgroups at electroweak scale is the $\susix$ GUT \cite{hartanto1}. Unfortunately, the model is suffered from realizing the desired breaking patterns through  Higgs mechanism. It has been  concluded that the allowed Higgs multiplets in the model are not able to reproduce all particle spectrums within the  experimental bounds.

Nevertheless, it has been found that the symmetry breaking for $\susix$ GUT can be realized by introducing an extra dimension at the $\susix$ scale\cite{hartanto2}. This non-Higgs mechanism is adopting the Scherk-Schwarz mechanism to dynamically break the symmetry induced by the orbifold of extra dimension \cite{nomura1, nomura2, quiros}. The effect on compactifying the extra dimension is considered to induce the Higgs bosons itself, and is known as the Higgs -- gauge boson unification \cite{nomura3}. Later, the generated Higgs bosons are utilized to break the subsequent symmetry to the electroweak scale\cite{hartanto2}. This paper is focused on investigating the proton decay within the model with the above-mentioned breaking mechanism. Because proton decay is the most severe constraint for any models beyond the SM.

The paper is organized as follows. First the symmetry breaking to $\suthr_C \otimes \suthr_H \otimes \uone_C$ due to dimensional reduction with orbifold $\sz$ is briefly reviewed. Subsequently it is shown that the adopted mechanisms in this case naturally suppress the proton decay since the leptoquark like interactions are not allowed at tree level as happened in the experimentally excluded conventional SU(5) GUT. Finally the paper is concluded with a short summary.

\section{5$-$dimensional $\susix$}
\label{sec:su6}

The 4-dimensional (4D) $\susix$ model is already given in our previous work \cite{hartanto1}, but in the present paper let us consider the extended model laying on 5D space-time. The 5D bulk in the model is divided into two branes due to the properties of orbifold breaking. One brane contains the 5D gauge bosons, $A_M$, while another particle contents live in the remaining 4D one. Each brane corresponds to the orbifold fixed point $y = 0$ and $\pi R$, and the parity transformation operator $Z_2$ is taken as,
\be
    Z_2^{(0)}=Z_2^{(1)}=\begin{pmatrix}
                          1 & 0 &   &    &    &   \\
                          0 & 1 &  &    &    &   \\
                            &  & 1 & 0  &    &   \\
                            &   & 0 & -1 &   &   \\
                            &   &   &  & -1 & 0 \\
                            &   &   &    &  0 & -1 \\
                        \end{pmatrix} \, .
\ee
This yields $U = Z_2^{(1)}Z_2^{(0)}=I_6$ is a unitary matrix.

Following the gauge-Higgs unification principle, one can define scalar boson sextet $\Phi$ in such a way that $A_M \equiv \Phi $ and $A_M^{\pm}(x,y)= \tilde{\Phi}_{\pm}(x,y)$. Here one may also define two vacuum expectation values (VEV's), $v$ and $v'$ at one fixed point $(y=0)$ as below,
\be
    v = \begin{pmatrix}
                     0 \\
                     0 \\
                    f_1 \\
                     0 \\
                     0 \\
                     0 \\
                   \end{pmatrix} \, , \quad \textrm{and} \quad
                   v' =\begin{pmatrix}
                     0 \\
                     0 \\
                     0 \\
                     0 \\
                     0 \\
                     f_2 \\
                   \end{pmatrix} \, .
\ee
Then, the $\susix$ Little Higgs in 4D world is found to be \cite{schmaltz,lhsg,hartanto3},
\bea
  \tilde{\Phi}^{(1)'}_+ & = & \frac{1}{\sqrt{\pi R}}
      \exp \left[ \frac{if_2}{ff_1}
                         \begin{pmatrix}
                           (0)_{3 \times 3} & \left( \begin{smallmatrix}
                                               \begin{smallmatrix}
                                                   0 & 0 \\
                                                   0 & 0 \\
                                                \end{smallmatrix}
                                                          & h \\
                                                 h'^{\dag} & 0 \\
                                               \end{smallmatrix} \right)
 \\
                           \left(\begin{smallmatrix}
                                               \begin{smallmatrix}
                                                   0 & 0 \\
                                                   0 & 0 \\
                                                \end{smallmatrix}
                                                          & h' \\
                                                 h^{\dag} & 0 \\
                                               \end{smallmatrix}\right)  & (0)_{3 \times 3} \\
                         \end{pmatrix} \right] 
         \begin{pmatrix}
                                        0 \\
                                        0 \\
                                        f_1 \\
                                        0 \\
                                        0 \\
                                        0 \\
                                      \end{pmatrix} \, , 
 \label{exp1} \\
  \tilde{\Phi}^{(2)'}_+ & = & \frac{1}{\sqrt{\pi R}}
    \exp \left[ -\frac{if_1}{ff_2}
                         \left(\begin{matrix}
                           (0)_{3 \times 3} & \left( \begin{smallmatrix}
                                               \begin{smallmatrix}
                                                   0 & 0 \\
                                                   0 & 0 \\
                                                \end{smallmatrix}
                                                          & h \\
                                                 h'^{\dag} & 0 \\
                                               \end{smallmatrix} \right)
 \\
                           \left(\begin{smallmatrix}
                                               \begin{smallmatrix}
                                                   0 & 0 \\
                                                   0 & 0 \\
                                                \end{smallmatrix}
                                                          & h' \\
                                                 h^{\dag} & 0 \\
                                               \end{smallmatrix}\right)  & (0)_{3 \times 3} \\
                         \end{matrix}\right) \right]
         \left(\begin{matrix}
                                        0 \\
                                        0 \\
                                        0 \\
                                        0 \\
                                        0 \\
                                        f_2 \\
                                      \end{matrix}\right)\, , 
 \label{exp2}
\eea

Now we are ready to proceed with the first symmetry breaking of the 5D $\susix$ GUT.

\section{Symmetry breaking : $\susix \rightarrow \suthr \otimes \suthr \otimes \uone$}
\label{sec:sb1}

According to the Scherk-Schwarz mechanism, the first breaking is performed by using the commutative and anti-commutative relations \cite{quiros},
\bea
  [Q',Z^{(0)}_{2}] & = &[\lambda^{a},Z^{(0)}_{2}]=0 \, ,\\
  \{Q,Z^{(0)}_{2}\} & = & \{\lambda^{\hat{a}},Z^{(0)}_{2}\}=0 \, .
\eea
Here $Q'=\lambda^{a}$ is the unbroken $\susix$ generators, while $Q'=\lambda^{\hat{a}}$ is the broken ones with $a = 1,2,\cdots,8, 27,28,\cdots,34, 35$ and $\hat{a}= 9,10,\cdots,26$. These induce the symmetry breaking,
\be
    \underbrace{\susix}_{5\mathrm{D}} \rightarrow \underbrace{\suthr \otimes \suthr \otimes \uone}_{4\mathrm{D}} \; ,
\ee
and the parity is splitted into : even parity for $A_{\mu}^{a}$ and $A_{y}^{\hat{a}}$; and odd parity for $A_{\mu}^{\hat{a}}$ and $A_{y}^{a}$. In this case the fermions are also even. Because the fields are transformed as follows,
\bea
  A_{\mu}(x,-y) & = & Z^{(0)}_{2}A_{\mu}(x,y) Z^{(0)}_{2} \, ,\\
  A_{y}(x,-y)   & = & Z^{(0)}_{2}A_{y}(x,y) Z^{(0)}_{2} \, ,\\
  \psi^{f}(x,-y)& = & \gamma_{5}Z^{(0)}_{2}\psi^{f}(x,y) \, ,
\eea
where $Z^{(0)}_{2}$ fulfills the consistency condition, $U Z^{(0)}_{2} U = Z^{(0)}_{2}$ \cite{nomura2}. Therefore, the remaining zero-modes at the low energy 4D effective theory are $A_{\mu}^{a(0)}$, $A_{y}^{a(0)}$, $\psi^{(0)}_{jR}$ ($j=1,2,3$) and $\psi^{(0)}_{jL}$ ($j=4,5,6$).

Applying the gauge-Higgs unification one obtains the scalar triplet bosons $\phi^{(i)}$ ($i=1,2$) which can be derived from the  upper and lower triplets of $\susix$ Little Higgs in Eqs. (\ref{exp1}) and (\ref{exp2}) \cite{hartanto3},
\bea
    \phi^{(1)} & = & \frac{1}{\sqrt{\pi R}} \ex^{\frac{if_2}{ff_1}\left(\begin{smallmatrix}
                                               \begin{smallmatrix}
                                                   0 & 0 \\
                                                   0 & 0 \\
                                                \end{smallmatrix}
                                                          & h' \\
                                                 h^{\dag} & 0 \\
                                               \end{smallmatrix}\right)}\left( \begin{matrix}
                                             0 \\
                                             0 \\
                                             f_1 \\
                                           \end{matrix}\right) \, , 
  \label{eq:p1j1} \\
    \phi^{(2)} & = & \frac{1}{\sqrt{\pi R}} \ex^{-\frac{if_1}{ff_2}\left(\begin{smallmatrix}
                                               \begin{smallmatrix}
                                                   0 & 0 \\
                                                   0 & 0 \\
                                                \end{smallmatrix}
                                                          & h \\
                                                 h'^{\dag} & 0 \\
                                               \end{smallmatrix}\right)}\left( \begin{matrix}
                                             0 \\
                                             0 \\
                                             f_2 \\
                                           \end{matrix}\right) \, .
  \label{eq:p1j2} 
\eea
Rescaling the VEV by a factor of $1/{\sqrt{\pi R}}$, and defining $(f'_1)^2+(f'_2)^2=(f')^2$ and $f'= {f}/{\sqrt{\pi R}}$, Eqs. 
(\ref{eq:p1j1}) and (\ref{eq:p1j2}) become,
\bea 
  A^{\hat{a}(1)}_{y}=\phi^{(1)} & = & \ex^{\frac{if'_2}{f'_1f'}\left(\begin{smallmatrix}
                                               \begin{smallmatrix}
                                                   0 & 0 \\
                                                   0 & 0 \\
                                                \end{smallmatrix}
                                                          & H' \\
                                                 H^{\dag} & 0 \\
                                               \end{smallmatrix}\right)}\left( \begin{matrix}
                                             0 \\
                                             0 \\
                                             f'_1 \\
                                           \end{matrix}\right) \, ,
\label{eq:jAa1}\\
A^{\hat{a}(2)}_{y}=\phi^{(2)} & = & \ex^{-\frac{if'_1}{f'_2f'}\left(\begin{smallmatrix}
                                               \begin{smallmatrix}
                                                   0 & 0 \\
                                                   0 & 0 \\
                                                \end{smallmatrix}
                                                          & H \\
                                                 H'^{\dag} & 0 \\
                                               \end{smallmatrix}\right)}\left( \begin{matrix}
                                             0 \\
                                             0 \\
                                             f'_2 \\
                                           \end{matrix}\right) \, ,
\label{eq:jAa2}
\eea
where $H = {h}/{\sqrt{\pi R}}$ and $H' = {h'}/{\sqrt{\pi R}}$. Eqs. (\ref{eq:jAa1}) and (\ref{eq:jAa2}) provide the scalar triplets required in 4D $\suthr \otimes \suthr$. The simplest little Higgs scenario is exactly reproduced by the special case  with $H = H'$ \cite{schmaltz,pbsm,lhr}.

Before moving forward, let us consider the particle assignment in the branes. The IR-brane at fixed point $y = 0$ contains all even particles : unbroken $\suthr$ gauge bosons, scalar triplet bosons and chiral fermions. Therefore the SM gauge bosons and fermions with even parity reside in IR-brane and the symmetry can accordingly labeled as usual : $\suthr_C \otimes \suthr_H \otimes \uone_{C_1/C_2}$. Consequently the massive $\suthr$ gauge bosons corresponding to the broken generators  $A^{\hat{a}}_{\mu}$ must reside in another UV-brane at fixed point $y = \pi R$ with off-diagonal $3 \times 3$ matrices. The  lower-left is basically the hermitian conjugate of the upper-right which justifies the  symmetry : $\suthr_L \otimes \suthr_R \otimes \uone_{C_2}$. Remind that $C_1$ and $C_2$ bosons appear from $\susix$ $C-$boson with generators \cite{hartanto1},
\be
    \lambda_{C_1}=\frac{1}{3}\sqrt{6}\begin{pmatrix}
                                      1 &  &  \\
                                       & 1 &  \\
                                       &  & 1 \\
                                    \end{pmatrix}, \, \, 
  \lambda_{C_2}=\frac{\sqrt{6}}{3}\begin{pmatrix}
                                      -1 &  &  \\
                                       & -1 &  \\
                                       &  & -1 \\
                                    \end{pmatrix} \, .
  \label{eq:cb}
\ee

Same with the odd particle $A^{\hat{a}}_{\mu}$, there is also another particle with odd-parity, that is $A^{a}_{y}$. This is a massless fifth-dimensional gauge boson, and under $Z_2$ symmetry fulfills the condition $ \oint dy A_{y}^{a}=0$. Moreover, the fifth-dimensional gauge invariance leads to $A^{a}_{y}=0$ which also satisfies the condition, and therefore it cannot stay in  UV-brane. Since fermions and scalar bosons live in IR-brane, it is obvious that UV-brane should contain only odd heavy gauge bosons from the broken $\susix$ and its subsequent $\suthr$ symmetries.

\section{Symmetry breaking : $\suthr_H \rightarrow \sutwo_L \otimes \uone_{B_1/B_2}$ and $\suthr_L \otimes \suthr_R \rightarrow \suthr_V$}
\label{sec:sb2}

The first breaking involves $B_1$ / $B_2$ boson that is generated from $\susix$ generators. Both are combined with $C_{1,2}$ bosons to form hypercharges for quarks and leptons through $\uone_{B_1/B_2} \otimes \uone_{C_1/C_2}\rightarrow \uone_{Y_q / Y_l}$. This can be achieved by putting $B_1 = 0$ and $B_2 = \lambda_{34}$.

The second breaking can be realized by bringing the little Higgs triplet,
\be \label{eq:lhmt}
    \phi^{(1)'}=f'_1 \ex^{\frac{i}{f'_1}\left(\begin{smallmatrix}
                                               \begin{smallmatrix}
                                                   0 & 0 \\
                                                   0 & 0 \\
                                                \end{smallmatrix}
                                                          & H' \\
                                                 H^{\dag} & 0 \\
                                               \end{smallmatrix}\right)}\left( \begin{matrix}
                                             1 & &  \\
                                              & 1 &  \\
                                              & & 1 \\
                                           \end{matrix}\right) \, ,
\ee
where the VEV is rewritten as ${f'}_1 \mathbf{1}_{3 \times 3}$ from $C_1$ boson in Eq. (\ref{eq:cb}) to achieve non-vanishing VEV and $H \sim H'$. From the results, there are two important properties : 1) the heavy gauge bosons have zero / integer charge and not fractional one like quark charges; 2) no tree-level interaction among heavy gauge bosons and SM particles due to opposite parity and the separated branes.

\section{Proton decay}
\label{sec:protondecay}

The main concern as developing the alternative GUT is the severe constraint from proton decay. In contrast to the present case, the original 4D SU(5) GUT contains leptoquarks. The exotic bosons interacts with the SM particles at tree level. This immediately provide too large contribution to the proton decay which later on exceeds the experimental bounds.

The sextet and decapentuplet in $\susix$  consist of the same quark (anti-quark) -- anti-lepton (lepton) for $\bar{6}$ and 6. On the other hand, for $\overline{15}$ and $15$ both consist of quark (anti-quark) -- anti-quark (quark) -- anti-lepton (lepton). The structures are the same as SU(5) except that there are additional heavy neutrinos $N$ ($N^c$)\cite{hartanto1}. Therefore, without any manipulation $\susix$ must be suffered from the same problem with SU(5).

Fortunately, as already discussed in the preceding section, the unique properties of fifth-dimensional orbifold might totally improve the situation. The symmetry breaking via Scherk-Schwarz mechanisms with orbifold breaking induces parity splitting such  that the unbroken gauge bosons $A^{a}_{\mu}$ and the broken fifth-dimensional component gauge bosons $A^{\hat{a}}_{y}$ with even parity reside in IR-brane $(y=0)$. While the broken gauge bosons $A^{\hat{a}}_{\mu}$ and the unbroken fifth-dimensional component gauge bosons $A^{a}_{y}$ with odd parity are projected out of IR-brane and reside in UV-brane $(y=\pi R)$. This property prevents the broken heavy gauge bosons to be leptoquark like bosons. Subsequently the tree level interactions among particles reside in both branes are not allowed. This guarantees that all corresponding processes in SU(5) contribute to the proton decay can not take place. Of course this fact also conserves the baryon number.

\section{Summary}
\label{sec:summary}

The symmetry breaking of $\susix$ GUT model with one extra  dimension through Scherk-Schwarz mechanism with orbifold $\sz$ breaking has been briefly introduced. The mechanism has succeeded in realizing two steps of breakings up to the SM scale, where the second breaking is realized by Higgs mechanism using little Higgs scalar bosons generated dynamically from the first breaking.

It has been shown that the above mentioned breaking mechanism avoids the emerging leptoquark bosons as happened in SU(5) GUT, and also prevents tree level interactions among particles reside in different branes.

\section*{Acknowledgments}

AH thanks the Group for Theoretical and Computational Physics, Research Center for Physics, Indonesian Institute of Sciences for warm hospitality during this work. The work of LTH is supported by the Riset Kompetitif LIPI 2010 under contract no  11.04/SK/KPPI/II/2010, while the work of FPZ is supported by Research KK-ITB 2010 and Hibah Kompetensi DIKTI 2010.

\bibliographystyle{ws-procs975x65}
\bibliography{protondecaySU6}

\end{document}